\documentclass[twocolumn,tighten,astrosymb,twocolappendix,times]{aastex631}

\begin{document}

\title{Comparison of Global \ion{H}{1} and H$\alpha$ Line Profiles in MaNGA Galaxy Pairs with FAST}

\author[0000-0003-3087-318X]{Gaoxiang Jin} 
\affiliation{Max Planck Institute for Astrophysics, Karl-Schwarzschild-Str. 1, D-85741, Garching, Germany}
\affiliation{Chinese Academy of Sciences South America Center for Astronomy (CASSACA), National Astronomical Observatories(NAOC),
20A Datun Road, Beijing 100012, China}

\author[0000-0002-7928-416X]{Y. Sophia Dai} 
\affiliation{Chinese Academy of Sciences South America Center for Astronomy (CASSACA), National Astronomical Observatories(NAOC),
20A Datun Road, Beijing 100012, China}
\correspondingauthor{Y. Sophia Dai}
\email{ydai@nao.cas.cn}

\author[0000-0003-0202-0534]{Cheng Cheng} 
\affiliation{Chinese Academy of Sciences South America Center for Astronomy (CASSACA), National Astronomical Observatories(NAOC),
20A Datun Road, Beijing 100012, China}

\author[0000-0002-1588-6700]{Cong Kevin Xu} 
\affiliation{Chinese Academy of Sciences South America Center for Astronomy (CASSACA), National Astronomical Observatories(NAOC),
20A Datun Road, Beijing 100012, China}

\author[0000-0001-6511-8745]{Jia-Sheng Huang} 
\affiliation{Chinese Academy of Sciences South America Center for Astronomy (CASSACA), National Astronomical Observatories(NAOC),
20A Datun Road, Beijing 100012, China}
\affiliation{Center for Astrophysics $|$ Harvard-Smithsonian, 60 Garden Street, Cambridge, MA 02138, USA}

\author[0000-0001-7218-7407]{Lihwai Lin} 
\affiliation{Institute of Astronomy \& Astrophysics, Academia Sinica, Taipei 10617, Taiwan}

\begin{abstract}

We present case studies comparing the global \ion{H}{1} and H$\alpha$ emission line profiles of six galaxy pairs.
The six pairs are selected to have different nuclear activities, with two hosting an active galactic nucleus,
and in different merging stages---two of each from pre-merging, merging, and post-merger stages.
We observe their global \ion{H}{1} spectra
with the Five-hundred-meter Aperture Spherical radio Telescope (FAST), 
achieving a noise level of $\sim$0.5\,mJy.
Five out of the six pair systems have secure detections of \ion{H}{1} emissions (signal-to-noise ratio $>$ 10).
The \ion{H}{1} fraction and star formation efficiency of the six pairs do not deviate from isolated galaxies.
For the \ion{H}{1} line profiles, common unique asymmetry is observed, indicating disturbances on the atomic gas from the galaxy interaction.
The global H$\alpha$ spectra of the merger systems are constructed from the optical integral field spectroscopic observations,
by integrating the flux in corresponding line-of-sight velocity bins.
The H$\alpha$ spectra tend to show multiple components in the pre-merger phase, 
and single component line profiles in the post-merger systems,
while all \ion{H}{1} spectra show single component line profiles regardless of merger stages.
The \ion{H}{1} and H$\alpha$ spectra show offsets in the central velocities,
which appear to decrease from $>100\rm \,km\,s^{-1}$ 
in the pre-merger pair
to $< 10\rm \,km\,s^{-1}$ in the post-mergers.
This trend is consistent with the scenario that, 
despite the significantly different distributions and kinematics of the atomic and ionized gases, 
the merging process may contribute to the 
mixing and eventually align various gas contents.

\end{abstract}


\section{Introduction} \label{sec:intro}

The interstellar and circumgalactic medium (ISM and CGM)
play important roles in the `baryon cycle' of the galaxy evolution.
Observationally, galaxies consume gas through star formation 
and the accretion of supermassive black holes (SMBHs).
The rapid consumption of gas indicates that galaxies have to obtain gas from the circumgalactic medium \citep[CGM, e.g.][]{2017ARA&A..55..389T}.
Also, the observed quenching of massive galaxies proves that there are mechanisms like ram pressure stripping \citep[e.g.][]{1972ApJ...176....1G} and AGN feedback \citep[e.g.][]{2012ARA&A..50..455F}, which can bring the gas away from the host galaxies and then stop the star formation.
In
simulations, galaxy-galaxy interactions and mergers
are the common fast ways for gas to flow into galaxies.
The gas inflow may enhance or trigger the star formation \citep[e.g.][]{1992ARA&A..30..705B} and the fast accretion of SMBHs (known as the active galactic nuclei, AGN) in merging systems \citep[e.g.][]{2008ApJS..175..356H}.

The enhancements of star formation are widely found 
among interacting and merging star-forming galaxies (SFGs),
based on their bluer color \citep[e.g.][]{2007ApJ...660L..51L}, stronger emission lines \citep[e.g.][]{2021ApJ...923..156D}, and luminous infrared emission \citep[e.g.][]{2011A&A...535A..60H}.
The strongest enhancements seem to occur in the central regions of SFGs \citep[e.g.][]{2019ApJ...881..119P}.
However, the merger impact on the star formation of individual galaxies is quite complex.
Previous works revealed that the merger-triggered star formation enhancement is related to several different parameters,
such as the merging stages \citep[e.g.][]{2019ApJ...881..119P},
nuclear properties \citep{2021ApJ...923....6J},
pair distance \citep[e.g.][]{2011MNRAS.412..591P},
mass ratio \citep[e.g.][]{2008AJ....135.1877E},
morphology \citep{2010ApJ...713..330X},
and bulge masses \citep{2022ApJS..261...34H}.
\citet{2021ApJ...923....6J} recently revealed that the
galaxy-galaxy interaction can enhance the star formation in SFG pairs,
but this effect is not significant in
narrow-line selected AGN-host galaxies or quiescent galaxies.
This difference may be directly related to the galaxy types, particularly their gas abundance,
since the narrow-line AGNs and quiescent galaxies are known to be more gas-deficient than SFGs \citep{2017ApJS..233...22S,2019MNRAS.482.5694E}.

For nearby galaxies, hydrogen gas is a major contributor to ISM and CGM.
The physical condition, spatial distribution, and kinematics of the hydrogen gas
offer information and precious probes for studying the physical and chemical processes during galaxies' evolution.
Based on the different temperatures and pressures,
hydrogen in the ISM and CGM exists mainly in three phases: the atomic, ionized, and molecular gas.
The neutral atomic gas (\ion{H}{1}) and ionized gas can be directly measured
through the fine structure emission line at 21\,cm and
the recombination lines at optical wavelengths, respectively.

\ion{H}{1}, as the most abundant and extended gas content in galaxies,
is the ideal indicator to study the kinematics of galaxy pair systems.
Simulations showed that the tidal forces during galaxy interaction
can trigger gas flows into the galaxy center \citep[e.g.][]{1996ApJ...471..115B},
while the resulted star formation or AGN activity
would quickly consume or blow out the gas.
Relevant and detailed observational evidence is still limited.
Marginal global \ion{H}{1} depletion ($\sim$15\%) is only found 
in tens of nearby major mergers \citep{2022ApJ...934..114Y},
while \citet{2018MNRAS.478.3447E} found a 0.3-0.6 dex enhancement of the \ion{H}{1} fraction.
To investigate the behavior of atomic gas during galaxy mergers,
more kinematic information is needed from \ion{H}{1} observations.
Interferometry studies on nearby interacting mergers found discrepancies between the tidal tails of \ion{H}{1} gas and stars \citep[e.g.][]{1996AJ....111..655H,2000AJ....119.1130H,2005ApJS..158....1I}.
But mapping the mass and velocity distribution of \ion{H}{1} in galaxies
is time-consuming and requires radio interferometry arrays,
making it expensive to build statistical galaxy merger samples with high resolution \ion{H}{1} maps.
An alternative approach
is to estimate the asymmetry from the shape of the global \ion{H}{1} line profile
through unresolved single-dish observations. 
For instance, \citet{2019MNRAS.484..582B} found that 
galaxy pairs tend to have more asymmetric global \ion{H}{1} line profiles.
\citet{2021MNRAS.504.1989W} and \citet{2022ApJ...929...15Z} suggested that although
pair and merger systems tend to have unique single-peaked line profiles,
there is no significant difference in the quantitative asymmetry distributions.
A recent case study also shows that the \ion{H}{1} content in a galaxy group
can extend to hundreds of kpcs from the group center \citep{2022Natur.610..461X}.
\citet{2023ApJ...956..148L} also found \ion{H}{1} stripping in interacting galaxy systems.
These complex \ion{H}{1} structures and kinematics in groups and pairs
can be the reason of the observed global line asymmetry.

Compared to \ion{H}{1}, ionized gas is a more direct tracer of star formation, 
since the hydrogen is mostly excited by the high energy photons from young stars.
These star-forming regions, also known as the \ion{H}{2} regions,
are the main contributors to the recombination lines in normal galaxies \citep{2006agna.book.....O}.
The global line profiles of strong recombination lines, such as H$\alpha$,
represent the global kinematics of the star formation component.
Conventionally, the optical H$\alpha$ spectra of nearby galaxies are typically observed
with narrow slit prisms or thin fibers,
thus either along a certain direction or limited to the central regions of a galaxy.
To obtain the global line profile of H$\alpha$,
optical spectra coverage is needed for the whole galaxy is needed.
Integral field unit (IFU) offers the opportunity to build up the optical spectra
of a galaxy.

\citet{2006ApJS..166..505A} presented a pioneer work of comparing the global \ion{H}{1} and H$\alpha$ line profiles of isolated face-on spirals.
The global H$\alpha$ line profiles are built and tested from the narrow band H$\alpha$ imaging and the IFU.
They found that most of the galaxies show agreement between the global \ion{H}{1} and H$\alpha$ line width, but the line shapes are significantly different, indicating possibly different locations or dynamics.
\citet{2009ApJ...700.1626A} and \citet{2023MNRAS.519.1452W} confirmed this conclusion in updated samples, and further investigated the origin of the asymmetry in both the global \ion{H}{1} and H$\alpha$ lines.
They suggested that the H$\alpha$ flux distribution typically dominates the asymmetry of the global H$\alpha$ spectra,
and most global \ion{H}{1} asymmetries trace disturbances in the outer regions of the host galaxies.

However, such studies are so far limited only to isolated, well-defined face-on galaxies. 
For galaxy mergers, the comparison between the atomic and ionized gas offers unique insight about the global gas kinematics and distributions during the merger event. 
It is suspected that the discrepancies between the asymmetry of H$\alpha$ and \ion{H}{1} result from strong perturbations in the galaxy scale, such as galaxy mergers  \citep{2023MNRAS.519.1452W}.

A sample of galaxy pairs and mergers with both IFU and \ion{H}{1} observations can be useful to examine whether galaxy interactions would induce different line profiles between
H$\alpha$ and \ion{H}{1} emissions.
One of the largest IFU surveys, 
Mapping the Nearby Galaxies at APO \citep[MaNGA,][]{2015ApJ...798....7B},
has observed $\sim$10000 nearby galaxies.
MaNGA observes the target galaxies out to 1.5 $r$-band effective radii ($R_e$),
meaning that the fiber bundles can cover most of the stellar and ionized gas component of the galaxies, or in pairs.
There is also a large sample of $\sim 1000$ merging galaxies observed by MaNGA \citep[][]{2019ApJ...881..119P,2021ApJ...923....6J}.
We take these advantages
and select a pair sample to
compare the global \ion{H}{1} and H$\alpha$ line profiles.

\ion{H}{1} line profile analysis needs high signal-to-noise ratio as well as enough velocity resolution.
For MaNGA galaxies at $z<0.05$,
\ion{H}{1}-MaNGA \citep{2019MNRAS.488.3396M} is the ongoing \ion{H}{1} follow-up survey. 
Its data include the proposed new observations using the Robert C. Byrd Green Bank Telescope (GBT), 
and the archive results from the Arecibo Legacy Fast ALFA survey \citep[ALFALFA,][]{2005AJ....130.2598G}, 
with the spectra root-mean-square (rms) of $\sim$1.5\,mJy and $\sim$3.5\,mJy, respectively (velocity resolution $\rm \sim 10\,km\,s^{-1}$).
The GBT observation can only reach the signal-to-noise ratio (S/N) of 3 on a $z=0.02$ galaxy with $\rm M_{H\,I}=9.4\,M_{\odot}$.
In this paper, we use the Five-hundred-meter Aperture Spherical radio Telescope \citep[FAST,][]{2006ScChG..49..129N} to observe our six pair systems
and to get deeper observations (rms\,$<$\,0.5\,mJy, $\Delta v = 10$$\rm \,km\,s^{-1}$, $\sim 3-10 \times$ deeper than \ion{H}{1}-MaNGA),
which is powerful and necessary to analyze the line profile shapes of the \ion{H}{1} spectra.
Additionally, the large beam size ($\sim$2.9 arcmin) of FAST \citep{2020RAA....20...64J} 
can offer us the global \ion{H}{1} line profiles tracing the atomic gas environment of the whole merger systems.
Along with the global $\rm H\alpha$ line profiles constructed from MaNGA,
here we present case studies to compare
the kinematics of the atomic gas and ionized gas in nearby merging systems.

The paper is constructed as follows:
In Section~\ref{sec:sample} we introduce the target selection and the data product.
In Section~\ref{sec:fast} and Appendix~\ref{ap:dr} we present our FAST data reduction procedure in detail.
Section~\ref{sect:analysis} is the scientific analysis,
in which we compare the FAST \ion{H}{1} spectra with the integrated MaNGA H$\alpha$ spectra for each galaxy pair system.
We measure the \ion{H}{1} fraction and the atomic gas SFE and compare them with other \ion{H}{1} surveys in Section~\ref{sec:hi-scale}.

Throughout this paper, the velocities are calculated by $c\times z$, where $c$ is the speed of light and $1+z=\nu_{0}/\nu=\lambda/\lambda_{0}$, 
and are then converted to the local standard of rest (LSR) frame.
We adopt a cosmology with $H_{0}=70\,\rm km\,s^{-1}\, Mpc^{-1}$, $\Omega_{m}=0.3$, and $\Omega_{\Lambda}=0.7$.
All stellar masses and star formation rates are based on the Kroupa initial mass function \citep{2001MNRAS.322..231K}.

\section{Target Selection and Sample Properties}
\label{sec:sample}

\subsection{MaNGA}
\label{subsec:manga}
MaNGA
is one of the main surveys of the fourth generation of the Sloan Digital Sky Survey \citep[SDSS-IV,][]{2015ApJ...798....7B},
which has obtained the IFU spectra for over 10\,000 nearby galaxies.
The field of views (FOVs) of the MaNGA science IFUs vary in diameter from 12\arcsec \ to 32\arcsec, 
which cover most of the stellar component of our targets.
The 2\arcsec \ fibers have a spatial resolution of $\sim$1\,kpc at $z$\,=\,0.03.
The spectral resolution ($\lambda / \Delta \lambda$) is about $\sim$2000 at the $\rm H\alpha$ wavelength \citep{2013AJ....146...32S}.
Our analysis in this paper is based on the latest public data release, MaNGA Product Lanuch 11 \citep[the same as SDSS DR17,][]{2022ApJS..259...35A}.

Our parent sample is
the MaNGA galaxy pair sample used in \citet{2021ApJ...923....6J}.
This parent sample includes 994 IFU-covered galaxies in pairs. 
Morphologically, the pair systems are visually classified into merger cases
from pre-merging (isolated) to the final coalescence (post-merger). 
The sample is classified into
narrow-line AGNs, composite galaxies, star-forming galaxies, and quiescent galaxies
based on the emission line ratio diagnostics.
Here we refer the readers to \citet{2021ApJ...923....6J} for more details about the pair selection, merger stage definition, and the AGN classification.

We use the redshifts and stellar masses in MaNGA's parent sample catalog, 
the NASA-Sloan-Atlas\footnote{NSA; M. Blanton; \url{http://www.nsatlas.org/}\label{footnote:nsa}}.
These redshifts are derived from the SDSS spectra, which are observed by single fibers targeting at the photometric center of the galaxies.
The stellar masses are calculated from the multiwavelength spectral energy distribution (SED) fitting \citep{2007AJ....133..734B}.
The photometries used for the fitting include UV bands from GALEX \citep[The Galaxy Evolution Explorer, ][]{2005ApJ...619L...1M}
and optical bands from SDSS.
The results based on elliptical Petrosian aperture photometries
are chosen to reduce the uncertainty due to irregular morphologies of galaxy pairs.
The global star formation rates are calculated by the attenuation-corrected $\rm H\alpha$ luminosities,
following \citet{2012ARA&A..50..531K}:
\begin{equation}
   {\rm \log(\frac{SFR}{M_{\odot}\,yr^{-1}}) = \log (\frac{L_{H\alpha}}{erg\,s^{-1}}) - 41.27}.
   \label{eq:sflaw}
 \end{equation}
The $\rm L_{H\alpha}$ is based on the integrated spectra in MaNGA's FOV
and corrected for attenuation by assuming an intrinsic $\rm H\alpha/H\beta$=2.86 \citep[Case B recombination, ][]{2006agna.book.....O}
and the reddening curve from \citet{1989ApJ...345..245C}:
$L_{\rm H\alpha} = L_{\rm H\alpha, obs}\times {\rm [(H\alpha/H\beta)_{obs}/2.86]^{2.36}}$.
We use the results from MaNGA Data Analysis Pipeline \citep[DAP, ][]{2019AJ....158..160B,2019AJ....158..231W} for the $\rm H\alpha$ flux and velocity measurements.

\subsection{Target selection}
\label{subsec:selection}
For the FAST observations, we selected six pair systems to represent different merging stages
and pair types.
We restrict the pairs to have spectroscopic redshifts of $z<$\,0.03,
in order to reach high S/N and avoid radio frequency interference (RFI) at lower frequencies \citep{2020RAA....20...64J}.
Within 3 arcmins, there are no galaxies at similar redshift (i.e., $\Delta v < 2000 \rm \,km\,s^{-1}$),
which ensures that the pairs are physically isolated in the FAST central beam and the \ion{H}{1} observations are not contaminated
by nearby sources.
To cover different galaxy pairs along the merger sequence and compare the differences,
we select two pairs in each merger stage:
pre-merging stage (weak or no distortion),
merging stage (strong distortion),
and post-merger stage (coalescenced mergers).
There are two AGN host galaxies in the merging and post-merger stages, respectively.
This selection is made to enable the comparison
of the \ion{H}{1} line profiles for AGN pairs and non-AGN pairs.

\begin{figure*}
   \plotone{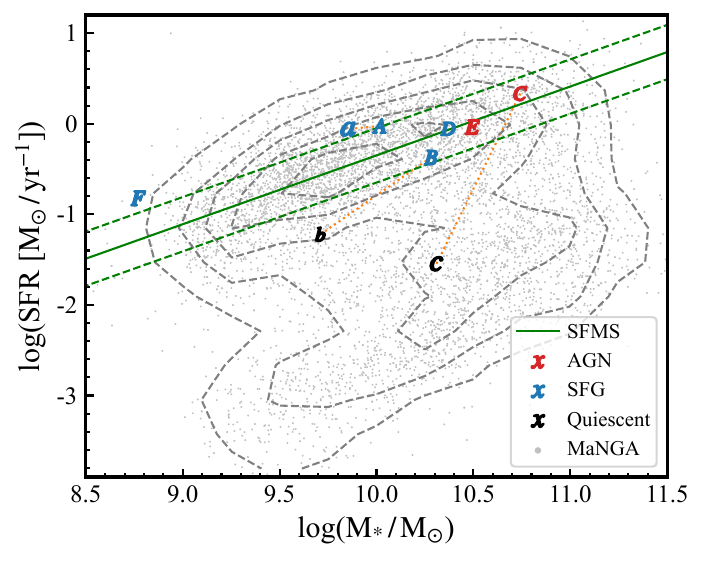}
   \caption{The stellar masses and SFRs of the galaxies in our six pair systems, 
   overlaid on the overall SFR-$M_*$ distribution (grey dots and contour) of all MaNGA galaxies at $z<0.05$. 
   The star formation main sequence (SFMS) as well as the typical scatter ($\pm\,0.3\,$ dex)
   are taken 
   from \citet{2015ApJ...801L..29R} and shown in green lines. 
   If separable, the stellar mass and SFR of the two member galaxies 
   in a pair are plotted individually and connected by the orange dotted lines.
   Galaxies in the same pair share the same symbols, and are color-coded 
   by their BPT types.
   \label{fig:sfms}}
   \end{figure*}

The MaNGA Plate-IFU numbers, sky coordinates, redshifts, AGN classification results, global stellar masses and SFRs, and merger stages of the six pair systems are summarized in Table~\ref{tab:sample}.
Fig.~\ref{fig:sfms} shows the global stellar masses and SFRs of the member galaxies in our pairs
or the pair systems if not separable.
The two AGNs, as well as the SFGs in the pairs, 
all lie along the star formation main sequence (SFMS) from \citet{2015ApJ...801L..29R}.

\begin{deluxetable*}{ccccccccc}
\tablecaption{Global properties of the 6 pair systems observed by FAST\label{tab:sample}}
\tablewidth{0pt}
\tablehead{
\colhead{Pair ID} & \colhead{Plate-IFU} & \colhead{R.A.} & \colhead{Decl.} & \colhead{Optical $z$} & \colhead{BPT Class} &
\colhead{$M_{*}$} & \colhead{SFR} & \colhead{Merger Case} \\
\colhead{} & \colhead{} & \colhead{deg} & \colhead{deg} & \colhead{} & \colhead{} &
\colhead{log($M_{\odot}$)} & \colhead{$M_{\odot}\, \rm yr^{-1}$} & \colhead{}
}
\decimalcolnumbers
\startdata
 A & 9194-12701 & 46.5605 & -0.3444 & 0.0287 & SFG & 10.01 & 0.94 & Pre-merging\\
 a & 9194-12701 & 46.5574 & -0.3416 & 0.0294 & SFG & 9.85 & 0.87 &\\
 \cline{1-9}
 B & 8254-12701 & 161.1697 & 44.0661 & 0.0258 & SFG & 10.28 & 0.36 & Pre-merging\\
 b & N/A        & 161.1447 & 44.0413 & 0.0251 & Quiescent & 9.71 & 0.05 &\\
 \cline{1-9}
 C & 8978-12705 & 249.5586 & 41.9388 & 0.0286 & AGN & 10.74 & 2.15 & Merging\\
 c & 8601-1902 & 249.5576 & 41.9311 & 0.0280 & Quiescent & 10.37 & 0.02 & \\
 \cline{1-9}
 D & 8241-12705 & 127.6320 & 18.2061 & 0.0269 & SFG & 10.37 & 0.88 & Merging\\
 \cline{1-9}
 E & 9507-12704 & 129.6001 & 25.7545 & 0.0182 & AGN & 10.49 & 0.92 & Post-merger\\
 \cline{1-9}
 F & 8725-9101  & 126.8250 & 46.0326 & 0.0073 & SFG & 8.77 & 0.15 & Post-merger\\
\enddata
\tablecomments{(1): Galaxies in the same pair are represented in the same letter but in different cases, with the capital letter denoting the more massive member galaxy.
(2): The MaNGA Plate-IFU number of the galaxies. 
(3)\&(4): The sky coordinates of the galaxies.
(5): Redshifts derived from SDSS optical spectra. 
(6): BPT classification for the galaxy centers using MaNGA spectra. 
(7)\&(8): Global stellar masses and SFRs of the galaxies (or the whole pair systems if the members are not separable). 
(9): The merger stages. }
\end{deluxetable*}

\section{FAST Observation and Data Reduction}
\label{sec:fast}

\subsection{Observational settings}
\label{subsec:setting}
The FAST observations were carried out in the latter half of 2021 during
the shared-risk period (proposal ID: PT2020-0160).
We used the L-band and the central beam (M01) of the FAST 19-beam receiver to observe the \ion{H}{1} 
emission lines.
The systematic performance,
such as aperture efficiency, pointing accuracy, and system temperature,
has been tested and discussed in \citet{2020RAA....20...64J}.
The frequency coverage of FAST L-band is from 1.05 GHz to 1.45 GHz,
with a channel resolution of 7.63 kHz 
(velocity resolution $\sim$1.6 km/s at $z\sim$0.02).
The average half-power beamwidth of the central beam is $\sim$2.9 arcmin.
Since the beamsize is $\sim$80 kpc in physical size at the targeted redshift,
much larger than the sizes of our pair systems ($<$1 arcmin),
the \ion{H}{1} content outside the central beam is not considered.

We adopt the ON-OFF observation mode, switching every five minutes.
The ON point of beam M01 is set at the galaxy pair,
while the OFF pointings are set at background skies tens of arcmins away,
chosen to have no galaxies at similar redshift.
Thus, the OFF pointings can serve as the approximate baselines at the target frequencies.
The high-power calibration noise is injected
during the first 20 seconds of each ON/OFF observation.

\subsection{Data reduction}
\label{subsec:reduction}
At the time of observation, FAST was still improving its performance especially the RFI issues
and did not have a finalized data reduction pipeline. 
We adopted the instrument parameters provided by \citet{2020A&A...638L..14C}, 
and built our own pipeline for our observation settings to reduce the data.
Our detailed 4-step data reduction procedure can be found in Appendix~\ref{ap:dr},
and is briefly summarized below. 
We first manually identify and remove the time-domain RFIs, 
and then perform the temperature calibrations for each exposure,
the standing waves baseline removal is carried out for the ON minus OFF spectra,
before converting to flux unit, and smoothing the stacked spectra
to measure the \ion{H}{1} emissions. 
The observation settings and spectral properties are listed in Table~\ref{tab:obs}.
The integration time ($t_{\rm Intn}$) is the `real' ON duration after removing the time affected by RFIs.

After the stacked \ion{H}{1} spectra have been obtained,
we calculate the \ion{H}{1} mass using 
the relation as firstly derived by \citet{1975gaun.book..309R}.
Assuming that the \ion{H}{1} emission is optically thin and unresolved by the single-dish telescope beam,
the \ion{H}{1} mass can be calculated as
\begin{equation}
\frac{M_{\rm H\,I}}{M_{\sun}} = \frac{2.36\times10^5}{(1+z)^{2}} (\frac{D}{\rm Mpc})^2 \int S(v)dv \
,
\label{eq:himass}
\end{equation}
where $S(v)$ is the line flux density\footnote{Here Jy and Jy$\rm \, beam^{-1}$ are the same for the flux density since all the sources are unresolved.} in $\rm Jy$,
and $D$ is the luminosity distance of the galaxy.
For non-detections we assume a line-width of 300\,$\rm km\,s^{-1}$ and integrate the rms
as the upper limit of the \ion{H}{1} mass.

The measured and calculated \ion{H}{1} properties are listed in Table~\ref{tab:result}.
The signal-to-noise ratio (SNR) is calculated by the ratio of the peak flux (after Hanning-smooth) and the rms.
We also compare the sensitivity of FAST with other telescopes in Appendix~\ref{ap:sensitivity}.
The central \ion{H}{1} velocities ($v_{\rm center}$) are
calculated as the mean value of the right and left wing velocities,
which are defined as the velocities at the 50\% of the peak 
flux for both sides of the emission line. 
The wing velocity widths ($W_{\rm P50}$)  are defined as the width of the HI line measured at 50\% of the peak between both sides, the same as in the \ion{H}{1}-MaNGA survey \citep{2019MNRAS.488.3396M}.
Values of the 50\% cumulative flux velocities ($v_{\rm c50,H\alpha}$, $v_{\rm c50,HI}$), as shown in Figure $\ref{fig:hahipre}$, Figure $\ref{fig:hahimid}$,  
and Figure $\ref{fig:hahipost}$, 
while $\Delta{v_{\rm c50}}$ 
is the difference between $v_{\rm c50,HI}$ 
and $v_{\rm c50,H\alpha}$,
and will be discussed in Section~\ref{sec:comparison}.

\begin{figure*}
   \centering
   \includegraphics[width=\textwidth, angle=0]{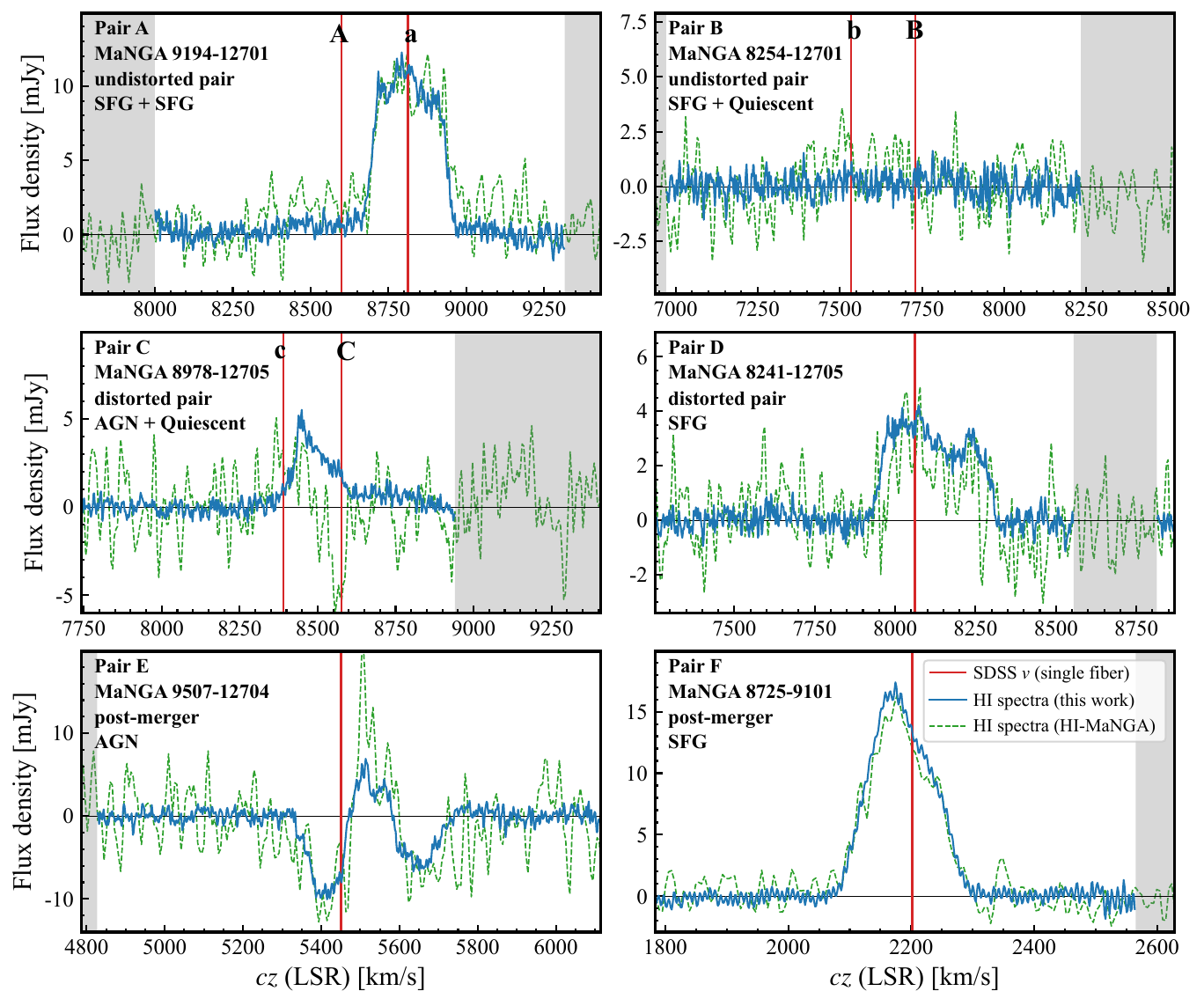}
   \caption{The \ion{H}{1} spectra of our six pair systems. 
   In blue are our FAST results with a velocity resolution of about 3.3$\rm \,km\,s^{-1}$.
   Red solid lines mark the optical velocities from SDSS single fiber spectra of the pair member(s),
   measured at the optical photometric center(s).
   The redshifts of each member in Pair A, B, C are also labeled by their pair ID.
   The frequencies affected by severe frequency-domain RFIs are masked and shown as gray shaded regions same as the yellow regions in Fig.~\ref{fig:pipeline}.
   For comparison, the archival spectra from the \ion{H}{1}-MaNGA survey \citep{2019MNRAS.488.3396M} are plotted in green dashed lines. The velocity resolution is $\sim$10 $\rm km\,s^{-1}$ with a typical on-target integration time of  $\sim$1000\,s. FAST observations show significant improvement in both the spectral resolution and the signal-to-noise ratio.}
   \label{fig:all}
   \end{figure*}

\begin{deluxetable*}{cccccc}
\tablecaption{Observational settings} \label{tab:obs}
\tablewidth{0pt}
\tablehead{
\colhead{Pair ID} & \colhead{ON R.A.} & \colhead{ON Decl.} & \colhead{OFF R.A.} & \colhead{OFF Decl.} & \colhead{$t_{\rm Int}$} \\
\colhead{-} & \colhead{-} & \colhead{-} & \colhead{-} & \colhead{-} & \colhead{s} 
}
\decimalcolnumbers
\startdata
    Pair A & 03:06:14.39 & -00:20:37.9 & 03:04:33.28 & -00:31:06.2 & 1482.9  \\ 
    Pair B & 10:44:40.76 & +44:03:59.5 & 10:43:41.05 & +44:04:31.0 & 599.4  \\
    Pair C & 16:38:14.44 & +41:56:33.2 & 16:41:28.69 & +42:00:00.6 & 2018.2  \\
    Pair D & 08:30:31.53 & +18:12:16.1 & 08:29:38.14 & +18:24:29.9 & 2377.5  \\
    Pair E & 08:38:24.12 & +25:45:15.0 & 08:37:36.08 & +26:07:56.5 & 299.0  \\
    Pair F & 08:27:18.32 & +46:02:09.8 & 08:27:18.04 & +46:05:01.1 & 2400.0  \\
\enddata
\end{deluxetable*}

\begin{deluxetable*}{cccccccccc}
\tablecaption{Observational results} \label{tab:result}
\tablewidth{0pt}
\tablehead{
\colhead{Pair ID} &
\colhead{$f_{\rm peak}$} & \colhead{rms} & \colhead{SNR} &
\colhead{log($M_{\rm H\,I}$)} &
\colhead{$v_{\rm center}$} & 
\colhead{$W_{\rm P50}$} & 
\colhead{$v_{\rm c50,H\alpha}$} & 
\colhead{$v_{\rm c50,HI}$} & 
\colhead{$\Delta v_{\rm c50}$} 
\\
\colhead{-} & 
\colhead{mJy} & 
\colhead{mJy} & 
\colhead{-} &
\colhead{log($M_{\odot}$)} & 
\colhead{$\rm km\, s^{-1}$} & 
\colhead{$\rm km\, s^{-1}$} & 
\colhead{$\rm km\, s^{-1}$} & 
\colhead{$\rm km\, s^{-1}$} & 
\colhead{$\rm km\, s^{-1}$}
}
\decimalcolnumbers
\startdata
    Pair A & 12.26 & 0.45 & 24.6 & 9.96 & 8816 & 164 & 8666 & 8803 & 138\\ 
    Pair B & 1.45 & 0.50 & - & $<$8.26 & - & - & - & - & -\\
    Pair C & 5.51 & 0.34 & 16.2 & 9.42 & 8518 & 142 & 8465 & 8503 & 38\\
    Pair D & 4.21 & 0.29 & 14.5 & 9.50 & 8136 & 209 & 8086 & 8100 & 14\\
    Pair E & 6.83 & 0.66 & 10.4 & - & - & - & - & - & -\\
    Pair F & 17.38 & 0.46 & 37.4 & 8.40 & 2195 & 110 & 2175 & 2181 & 6\\
\enddata
\end{deluxetable*}

\section{Line profile Comparison of \ion{H}{1} and H$\alpha$}\label{sect:analysis}
In this section, we directly compare the H$\alpha$ emission line profiles from MaNGA IFU data 
with the FAST \ion{H}{1} line profiles. Below we perform a
case-by-case study of the ionized gas and atomic gas properties in our galaxy mergers. 

The \ion{H}{1} line profiles of six pair systems are shown in Fig.~\ref{fig:all},
where the velocity resolution is smoothed to $\sim 3.3\, \rm km\,s^{-1}$.
Compared to the archival \ion{H}{1}-MaNGA survey observations (green dashed lines), our results show clear improvements in both SNR and spectral resolution.
For the H$\alpha$ flux density-velocity line profiles,
we use all valid spaxels within the whole MaNGA FOV and sum the H$\alpha$ fluxes in velocity bins of 20\,$\rm km\,s^{-1}$.
The velocity of each spaxel is calculated by $c\times z_{\rm H\alpha}$ and then converted to the LSR frame.
We chose the velocity bin of 20\,$\rm km\,s^{-1}$ because \citet{2021AJ....161...52L} concluded that the MaNGA spectra can provide reliable measurements of astrophysical velocity dispersions $\sigma \rm _{H\alpha}\sim 20 \, km\,s^{-1}$.

The long transition time ($\sim 10^{7}$ years) of the \ion{H}{1} line means that the \ion{H}{1} line profile represents the Doppler motion of the atomic gas \citep[e.g.][]{2012RPPh...75h6901P}, 
while the H$\alpha$ line is more likely to be affected by the other broadening mechanisms,
for example, the normal turbulent motions of ionized gas can increase the $\sigma \rm _{H\alpha}$ to $\rm \sim 25 \, km\,s^{-1}$ \citep[e.g.][]{2006ApJS..166..505A}, while some violent processes like shock and winds can increase the $\sigma \rm _{H\alpha}$ to $\rm \gtrsim 100 \,km\,s^{-1}$ \citep[e.g.][]{2017MNRAS.470.4974D}.
So, the H$\alpha$ fluxes and velocities used for global line profiles are calculated from the Gaussian fit on MaNGA spectra without the velocity dispersion ($\sigma \rm _{H\alpha}$) information.
This way the H$\alpha$ line profiles can approximately represent the global rotation of the ionized gas and are less affected by broadening from winds and shocks,
thus can be compared to the single-dish \ion{H}{1} line profiles directly.

For comparison, we then normalize the \ion{H}{1} and the H$\alpha$ global line profiles and plot them in the same LSR frame,
for a direct comparison of the atomic gas and ionized gas flux distribution along the line-of-sight velocities.
Here the H$\alpha$ flux is not corrected for attenuation, 
since the H$\beta$ line has low S/N in galaxy outskirts, making the attenuation correction unreliable.
In addition, we check the attenuation (e.g., $A_{V}$) calculated from the Balmer decrement, and find that the differences between the $A_{V}$ in the central 2.5\arcsec and the $A_{V}$ in $R_e$ are around 0.5, thus the attenuation correction should not affect our global line shape.

\begin{figure*}[]
   \centering
   \includegraphics[width=\textwidth]{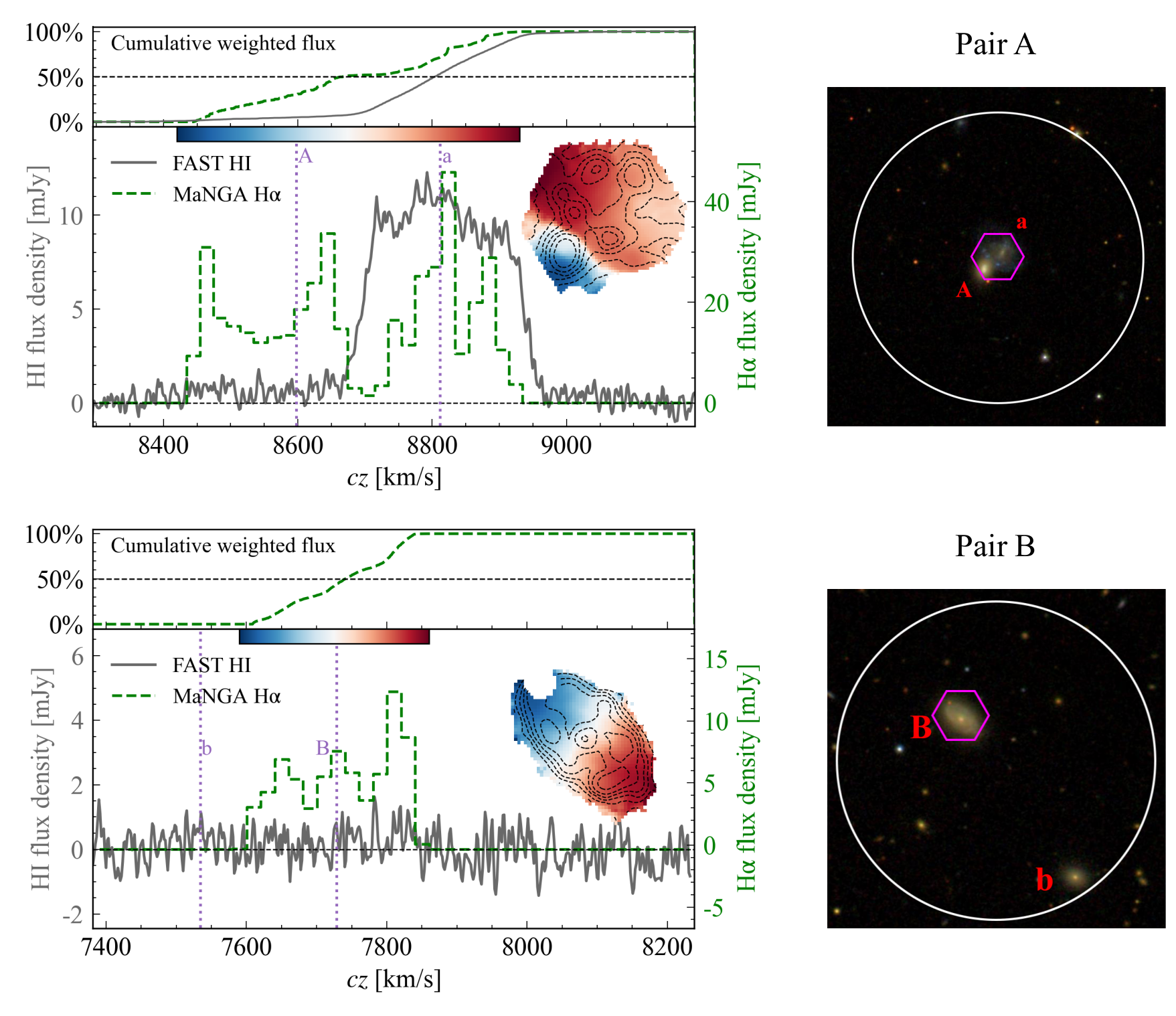}
   \caption{The comparison of global H$\alpha$-\ion{H}{1} spectra for pre-merging pars A and B. Pair B is not detected in \ion{H}{1}.
   The right panel shows the SDSS-$gri$ color image, 
   in which the magenta hexagon represents the MaNGA FOV
   and the white circle represents the FAST central beam location.
   The IDs of pair members are labeled near the galaxy positions.
   In the left panels, the gray lines and the green dashed lines are the cumulated
    \ion{H}{1} and H$\alpha$ spectra, respectively
   The SDSS redshifts are labeled in purple dashed lines, representing the optical velocities at the nuclear positions.
   The H$\alpha$ velocity maps from MaNGA are plotted in the panel, 
   color coded by the line-of-sight velocities, with color bars and corresponding velocity ranges 
   shown at the top of each spectra panel, along the velocity axis.
   Black dashed contours in the velocity maps are the H$\alpha$ flux maps.
   We also plot the cumulative spectra in the upper panels to distinguish the shapes between the two spectra.
   Their interceptions at the 50\% flux can be used to represent
   the line-of-sight velocity offsets for H$\alpha$ and \ion{H}{1}.
   \label{fig:hahipre}}
   \end{figure*}

Fig.~\ref{fig:hahipre}, \ref{fig:hahimid}, and \ref{fig:hahipost} show the comparison of H$\alpha$ and \ion{H}{1} line profiles of the pre-merging pairs, merging pairs, and post-mergers, respectively.
In these figures, the SDSS color images are shown in the right with the magenta hexagons and white circles representing the MaNGA FOV and FAST beam, respectively.
Pair members that are distinguishable are labeled in red text if available.
In the main panels, gray lines \ion{H}{1} spectra and green dashed lines are integrated H$\alpha$ spectra.
The H$\alpha$ velocity maps are plotted near the spectra, with H$\alpha$ flux maps shown in black dashed contours.
The colors are coded by the velocities,
while the color bars are matched to the velocity axis,
as an approximate link between the fluxes and the locations in the maps.
The upper panels are the cumulative fluxes of H$\alpha$ and \ion{H}{1}.
The steeper slope indicates that the fluxes are more concentrative in line-of-sight velocities.
The difference of the two 50\% flux velocities represents the line-of-sight offset of the ionized and atomic gas.

\subsection{Pair A (Fig.~\ref{fig:hahipre} upper panel)}
The system includes two SFGs with small projected separation (8.7 kpc) 
but a larger velocity offset (214.5 km\,s$^{-1}$).
There is no significant interaction feature, and the two galaxies show regular morphology and velocity maps,
indicating they are likely in the pre-merger stage.

\begin{figure}[]
   \centering
   \includegraphics[width=0.45\textwidth]{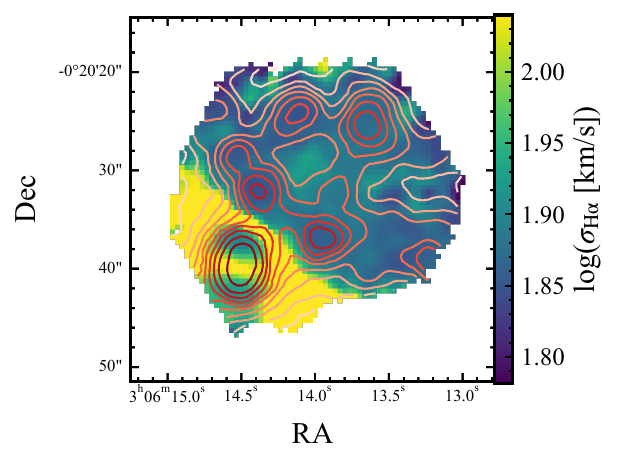}
   \caption{The H$\alpha$ velocity dispersion map of Pair A. The red contours are the H$\alpha$ flux density map. Despite the velocity differences, the two galaxies show clear different dispersions as well, indicating that the two members in the pair are kinematically distinct.}
   \label{fig:disp}
   \end{figure}

The MaNGA IFU covered most optical fluxes from both galaxies,
and the H$\alpha$ line profile shows two galaxy components, one rotating disk with the double horn structure, and one complex profile with possible differential structures.
The rotating disk component is more blueshifted and matched to the brighter galaxy in the lower right.
The fainter, bluer galaxy in upper right shows the complex line profile.
These two H$\alpha$ components are also visually distinct in the velocity map, as well as the optical image, stellar velocity map, and H$\alpha$ velocity dispersion map (Fig.~\ref{fig:disp}),
which indicates that these two galaxies are kinematically distinct in the line-of-sight direction.
This also confirms that they are in the pre-merging phase.

The \ion{H}{1} line profile, however, only shows one component, 
which only matches the velocities of the fainter galaxy.
At the velocity of the brighter galaxy, there are marginal signals weaker than 3$\sigma$.
The difference in the line profiles indicates that the \ion{H}{1} content of the pair may be associated with the fainter galaxy. It is not likely that the atomic gas has already mixed together in the pre-merging stage, because both the \ion{H}{1} spectra and the optical morphology do not show any signs of asymmetry or disturbance.

\subsection{Pair B (Fig.~\ref{fig:hahipre} lower panel)}
This system includes an SFG and a quiescent galaxy with large projected separation (57.2 kpc).
This is also a pre-merging pair system with no significant interaction features.
MaNGA covers the SFG and shows global star formation activity.

Although deeper than previous surveys such as \ion{H}{1}-MaNGA, 
we still detect no significant \ion{H}{1} signal. 
The $3\sigma\, \rm N_{HI}$ limit
is $\rm 2.8 \times\,10^{18}\, cm^{-2}$.
We conclude that this is
a relatively \ion{H}{1}-poor system including one SFG and one quiescent galaxy. 
We note that even though the system is not detected for \ion{H}{1}, the lower limit of the current atomic star formation efficiency (SFE) is still within the intrinsic scatter of the general SFE distribution (see also Fig.~\ref{fig:sfe}).

\begin{figure*}[]
   \centering
   \includegraphics[width=\textwidth]{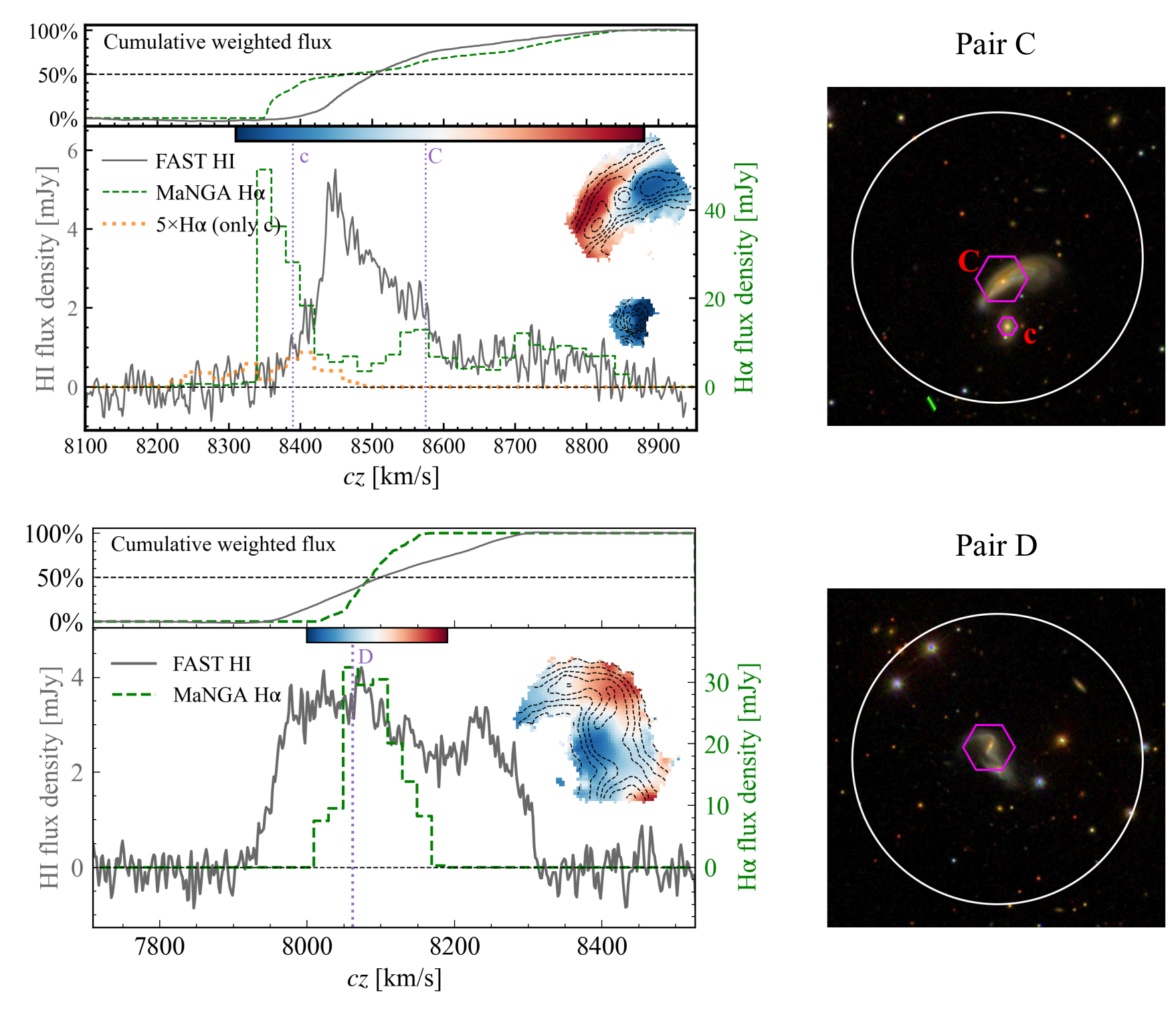}
   \caption{Same with \ref{fig:hahipre}. 
   Galaxy c only has marginal contribution to the overall H$\alpha$ line profile of the pair system.  In the upper panel, the H$\alpha$ profile from galaxy c is scaled up by a factor of five and plotted as the orange dotted line for a better demonstration. 
   The majority of the H$\alpha$ emission of the system is 
   from the upper right tail of the galaxy (bluer part). }
   \label{fig:hahimid}
   \end{figure*}

\subsection{Pair C (Fig.~\ref{fig:hahimid} upper panel)}
The system includes a narrow-line AGN with distorted morphology and a quiescent galaxy, 
covered by two MaNGA IFUs.
The clearly distorted morphology indicates that they are in the process of merging.

The H$\alpha$ flux concentrates near 8350\,km\,$\rm s^{-1}$, 
which is associated with a blueshifted star-forming region to the west of the central AGN.
The total velocity coverage of H$\alpha$, 
as indicated by the color bar in Fig.~\ref{fig:hahimid},
is about 500\,km\,$\rm s^{-1}$. 
The \ion{H}{1} line profile is asymmetric, extending over a range of $>$ 400\,km\,$\rm s^{-1}$,
with a long tail towards the red side of higher velocities.
The highest peak of \ion{H}{1} is $\sim$100\,km\,$\rm s^{-1}$ offset from the H$\alpha$ peak,
while there is a second peak of \ion{H}{1} at $\sim$8550\,km\,$\rm s^{-1}$,
which coincides with the central H$\alpha$ peak at the same velocity,
corresponding to the nuclear region of the AGN. 
We note that the H$\alpha$ from the southern galaxy (orange dotted line, amplified by 5 times, Fig.~\ref{fig:hahimid}) is weaker compared to the northern AGN.
The extended feature of the \ion{H}{1} line profile also 
matches well with
the H$\alpha$ fluxes at around 8600-8800\,km\,$\rm s^{-1}$.
The seemingly aligned velocity profile longwards of 8500\,km\,$\rm s^{-1}$ may 
indicate some intrinsic kinematic connection between the H$\alpha$ and \ion{H}{1} components.

The observational star formation law \citep[e.g,][]{2012ARA&A..50..531K} is well known to be a molecular rather than an atomic phenomenon \citep[e.g.][]{2008AJ....136.2846B},
which disfavors the correlation between H$\alpha$ and \ion{H}{1} in sub-galactic scales.
But during galaxy mergers, the large-scale disturbance can have significant impact on the kinematics of all phases of gas, for instance, redistributing the galaxy in the system towards similar kinematics,
which may eventually result in the line-of-sight velocity correlation between H$\alpha$ and \ion{H}{1}.
To confirm these implications, higher resolution radio interferometry observations are needed to confirm
this hypothesis. 

\subsection{Pair D (Fig.~\ref{fig:hahimid} lower panel)}
The system contains SFGs bridged together with clear tidal features.
The morphological connection of the two pair members
confirms that they are in the process of merging, 
probably already after the first encounter.

The wide \ion{H}{1} emission, with a width of $\sim$400\,km\,$\rm s^{-1}$,
fully covers and almost doubles the H$\alpha$ line width ($\sim$ 200 \,km\,$\rm s^{-1}$). 
We note that the MaNGA IFU does not fully cover the pair system.
The complete H$\alpha$ line width may be wider than observed
if the southern tail extends to high velocity.
Based on the H$\alpha$ velocity profile, 
the system is encountering volatile kinematics, with the two pair members 
and the tidal tails showing at several redshift/blueshift components.
Similar to Pair C, this value is at the high end of the 
reported \ion{H}{1} velocity width distributions in nearby galaxies
\citep[e.g.][]{2010MNRAS.403.1969Z,2022ApJS..261...21Y}.
Such high velocity width indicates that
either the merging activity strongly disturbed the atomic gas to spread in a wider velocity space,
or the \ion{H}{1} spectra consists of several gas components of overlapping velocities,
or a combination of both. 

\begin{figure*}[]
   \centering
   \includegraphics[width=\textwidth]{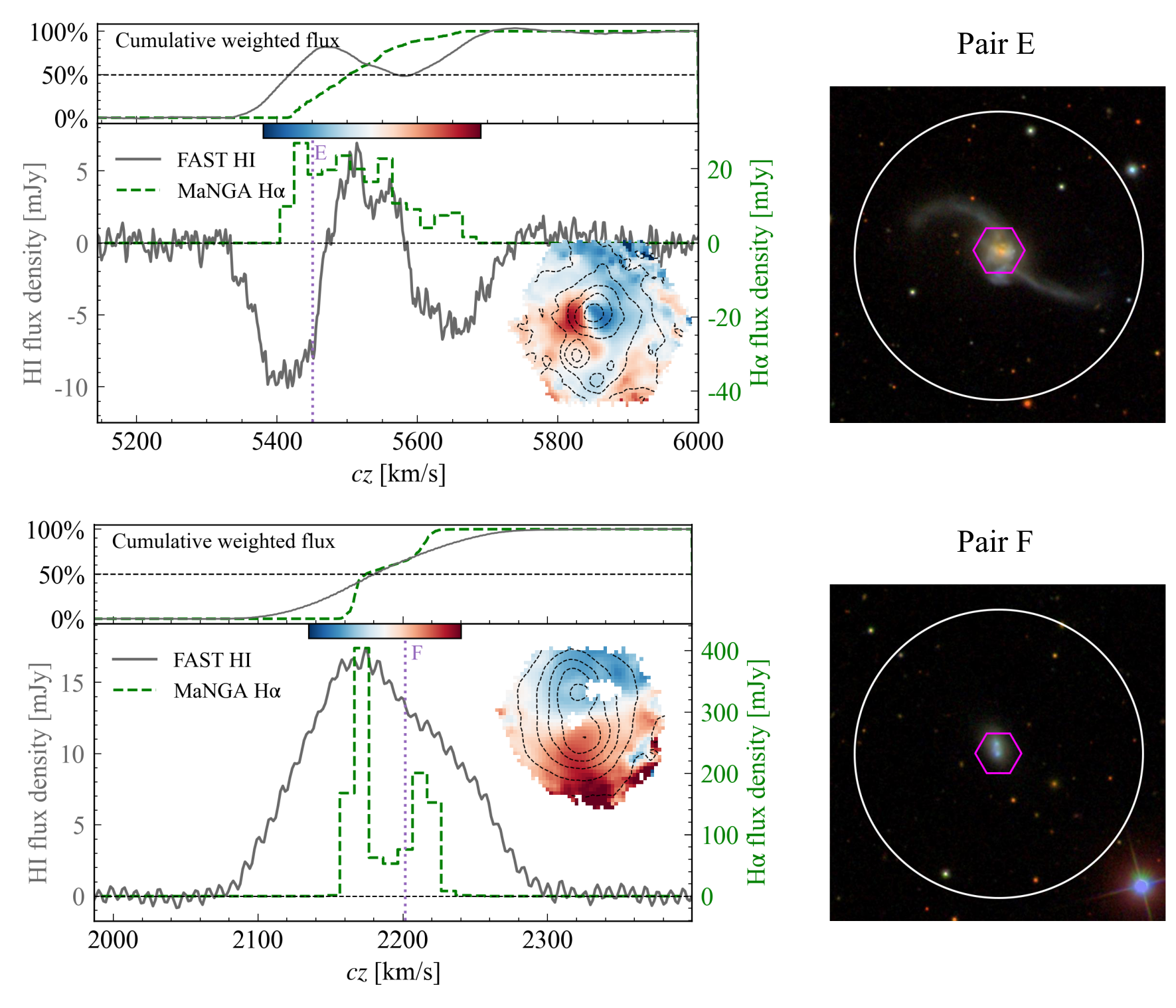}
   \caption{Same with \ref{fig:hahipre}, but for post-mergers E and F. Pair E show strong absorption features so that its cumulative \ion{H}{1} flux is not monotonically increasing.
   \label{fig:hahipost}}
   \end{figure*}

\subsection{Pair E (Fig.~\ref{fig:hahipost} upper panel)}
This is a clear post-merger system with two long, luminous tidal tails.
It is classified as a narrow line AGN.
MaNGA covers the central region, including the merger center and a luminous, blue star-forming region.

The H$\alpha$ components with the highest velocity component are close to the galaxy nucleus, 
where $\sigma \rm _{H\alpha}$ also reaches $\rm \sim 300\,km\,s^{-1}$ ($\sigma \rm _{*}$ is only $\rm \sim 180\,km\,s^{-1}$),
indicating a nucleus-driven line broadening.
We also note that the long tails are not covered by MaNGA,
the inclusion of the two tails may further widen the H$\alpha$ line profile.
The \ion{H}{1} spectra show two strong, broad absorption features.
Given the broad line width of the absorption features in the \ion{H}{1} line profile ($\rm \sim 400 \ km\,s^{-1}$), 
the absorption is possibly associated with the atomic gas related to the AGN and/or merger activities.

Follow-up high-resolution imaging and molecular gas observations are needed to identify the location and origin of the absorption feature.
We note that this \ion{H}{1} line profile shape, a central emission component with two strong absorption features at both wings, is rare and unique from most well-defined \ion{H}{1} absorption galaxies \citep[e.g.][]{2015A&A...575A..44G}.
The data reduction procedures are shown in Fig.~\ref{fig:pipeline} (c) and (d), and these features are not from the baseline fitting.
The emission and absorption features in the spectrum from FAST observation are consistent with those of the ALFALFA archival observations,
and the FAST spectrum shows a higher signal-to-noise ratio.
We will discuss this object in more detail in an upcoming paper (Dai et al., in preparation).

\subsection{Pair F (Fig.~\ref{fig:hahipost} lower panel)}
This is a low-mass post-merger system with
two separable nuclei in the optical image, covered in one MaNGA IFU.
The two nuclei are both classified as SFGs by the emission line diagnostics.

We can clearly see two H$\alpha$ components
from each galaxy in the flux map (top right corner, contours).
The northern nucleus is slightly blueshifted while the southern nucleus is redshifted from the SDSS reported redshift.
The H$\alpha$ flux map suggests two ionized gas cores around the optical center,
and they are not yet fully merged.
Thus, although the double-peaked shape of the global H$\alpha$ spectra could be created by a global rotation, it could also be explained by a mixture of two compact ionized gas cores with different line-of-sight velocities.

Instead of a double-horn shape for typical rotational atomic disks, 
the \ion{H}{1} line profile is centrally peaked.  
At the velocity of the southern nuclei ($\sim$2210\,km\,$\rm s^{-1}$),
the \ion{H}{1} also shows an excess hump, likely associated with the ionized gas. 
The line width of \ion{H}{1} (225$\rm \,km\,s^{-1}$) is three times wider 
than the H$\alpha$ ($\sim$70$\rm \,km\,s^{-1}$, peak-to-peak), 
confirming the common understanding that the atomic gas
is more spread out into the vicinity environment of the galaxy,
than the ionized gas that is often bound inside the galaxy \citep{1997ApJ...490..493N}.
We note that the limited MaNGA IFU size for this system may cause the H$\alpha$ line width to be underestimated,
but the H$\alpha$ velocity map already shows flattening in the northern and southern region,
indicating that the H$\alpha$ line is not likely to become as wide as the \ion{H}{1} line even with enough IFU coverage.

\begin{figure*}
   \centering
   \includegraphics[width=0.65\textwidth]{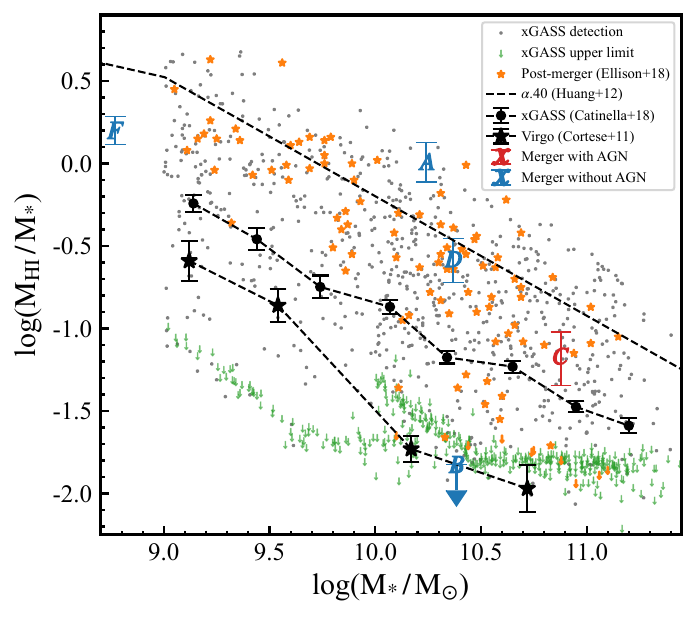}
   \caption{\ion{H}{1} fraction versus stellar mass. We compare the pairs in this paper with statistic samples in different gas environments from large \ion{H}{1} surveys, namely: \ion{H}{1}-rich sample from ALFALFA survey \citep{2012ApJ...756..113H}; \ion{H}{1}-normal sample from xGASS \citep{2018MNRAS.476..875C}; 
   and \ion{H}{1}-poor sample from the Virgo cluster \citep{2011MNRAS.415.1797C}.
   The pair with AGN (C) lies along the \ion{H}{1}-rich to \ion{H}{1}-normal regions,
   while the pairs without AGNs (A, B, D, F) show a wider range of \ion{H}{1} fractions,
   similar to the post merger sample (orange stars) in \citet{2018MNRAS.478.3447E}.}
   \label{fig:fhi}
   \end{figure*}

\subsection{Summary of the comparisons between the global \ion{H}{1} and H$\alpha$}
\label{sec:comparison}
In summary, the \ion{H}{1} line profiles typically show one continuous component
of linewidth between $\sim$220 to 400$\rm \,km\,s^{-1}$ in our sample. 
Wider \ion{H}{1} line profile ($\rm \sim\,400\,km\,s^{-1}$)
is found in pairs undergoing merging or just merged with distorted morphology (C, D, E, F),
indicating that the merging process directly affects the atomic gas in their environment. 
The H$\alpha$ line profiles, despite the lower velocity resolution,
often show velocity components peaking at one member galaxy or the center of the two member galaxies (A, C, D, F).
In most cases, the \ion{H}{1} velocity span covers both members, 
except in Pair A, where the HI profile is better aligned with one member galaxy. This SFG-SFG pair seems to be in the pre-merger case,
when the galaxy interaction may not have started yet.

In isolated galaxies, the atomic and ionized gas have been found to share similar kinematics, and the line shape difference could be due to the clumpier H$\alpha$ distributions, as reported before in \citet{2009ApJ...700.1626A}.
Considering that the volatile merging process should affect the gas kinematics in the system, regardless of the gas scale or type \citep[e.g.][]{,2013AJ....145...34S},
it is possible that merging would eventually align the atomic and ionized gases.

From a global point of view, we notice that the 50\% cumulative flux velocities
are generally different between the
\ion{H}{1}
and H$\alpha$ emissions (Table~\ref{tab:result}).
We note that given the different scales, distributions, and strengths of the ionized and atomic gases, 
it is normal that the widths, spans, and absolute values of the
\ion{H}{1} and H$\alpha$ emission line profiles 
are different. 
Therefore, we use $\Delta\,v_{\rm c50}$,
the differences between the bulk central velocities
of the \ion{H}{1} and H$\alpha$ emissions,
as a proxy for the alignment level between the two gas contents. 
Lower $\Delta\,v_{\rm c50}$ values indicate 
better alignment, and vice versa.
We notice that in the pre-merging case, Pair A, $\Delta\,v_{\rm c50}$ is $\sim 140\,\rm \,km\,s^{-1}$, though we also note that one ionized gas spectrum does not have \ion{H}{1} emission associated with it. 
As we move towards later merging cases, this value decreases from a few tens of $\rm km\,s^{-1}$ in merging cases (Pair C, D) to $<$10 $\rm km\,s^{-1}$ in the post-merger case (Pair F).

Despite our small sample size,
this trend suggests that the atomic and ionized gas
tend to share similar kinematic centers toward later
merging stages.
This is consistent with the scenario that
the merging process would eventually align and settle the different gas contents down to a common central velocity, 
possibly as a manifestation of gas concentration.

The overall line profiles of \ion{H}{1} and H$\alpha$ are generally different in all pairs in our sample,
except for Pair C, in which part of the spectra show some level of overlap. 
Our results are in general agreement with previous studies
that illustrated the environment impact on gases of different phases.
For the atomic gas, galaxy merger or a dense environment can increase the asymmetry of the spatial distributions \citep{2006MNRAS.369.1849A,2007MNRAS.378..276A},
which has also been confirmed in ionized gases
by several spatially resolved studies \citep[e.g.][]{2020ApJ...892L..20F,2023PASA...40...60B}.

\section{\ion{H}{1} scaling relations}\label{sec:hi-scale}
In this section, we investigate the \ion{H}{1} mass properties of the six pairs, and compare them to the scaling relations derived from large samples of field galaxies. 
We note that because the \ion{H}{1} observations do not resolve the pair members,  the \ion{H}{1} mass, stellar mass, and SFR used in this section are the integrated values for pair system as a whole.

The \ion{H}{1} fraction, defined as log($M_{\rm H I}$/$M_{*}$),
is used to estimate the atomic gas abundance of a galaxy.
The \ion{H}{1} fraction has been widely reported
to have a negative correlation with the stellar mass \citep[see][, for a review]{2022ARA&A..60..319S},
also known as the \ion{H}{1} scaling relation.
However, the \ion{H}{1} scaling relation is found to depend on the survey depth of the derived sample,
therefore with different values in the literature.
In Fig.~\ref{fig:fhi}, we compare our mergers with the extended GALEX Arecibo SDSS Survey \citep[xGASS,][]{2018MNRAS.476..875C} representative sample (gray dots and green upper-limits), as well as the scaling relation derived from several other surveys.
The dashed line is the scaling relation from the 40\% ALFALFA survey \citep{2012ApJ...756..113H}, 
which is mostly a local \ion{H}{1}-rich sample.
The dashed line with solid circles and errors is
from a deeper \ion{H}{1} survey, xGASS.
This scaling relation includes \ion{H}{1} non-detections, so can serve as a prediction of \ion{H}{1}-normal galaxies.
The dashed-star line is the relation from the galaxies in the Virgo cluster \citep{2011MNRAS.415.1797C},
representative of the \ion{H}{1}-poor galaxies in dense environments.

The pair systems with AGNs located near the average of the xGASS survey, indicating they are \ion{H}{1}-normal systems,
while the SFG mergers show a wide range of \ion{H}{1} fraction. 
The upper-limit SFG pair, Pair B, is a \ion{H}{1}-poor system.
Here we do not include the analysis for Pair E, since its \ion{H}{1} mass is not constrained due to strong absorption.

\begin{figure*}
   \centering
   \includegraphics[width=0.65\textwidth]{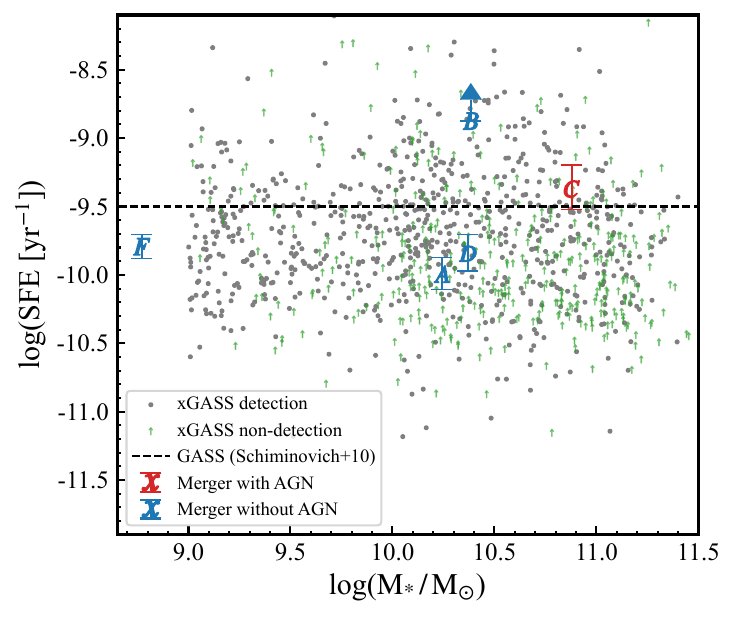}
   \caption{The atomic gas star-formation efficiency versus stellar mass. Compared to the xGASS survey (gray dots and green lower limits), the five \ion{H}{1}-detected pairs locate in normal SFE range. The non-detection pair B has relatively high star-formation efficiency. The black dashed line shows the average value of the xGASS survey sample.}
   \label{fig:sfe}
   \end{figure*}

The atomic gas star formation efficiency (SFE), defined as $\rm SFE=SFR/M_{HI}$,
is a proxy of how efficient the galaxy can convert the atomic gas to stars.
Previous \ion{H}{1} surveys have shown that the atomic gas SFE does not significantly evolve with the stellar mass,
and has an almost constant average value of $\rm \sim 10^{-9.5}\,yr^{-1}$ \citep{2010MNRAS.408..919S}.
Some studies also investigate the impact of merger on atomic gas SFE. 
For instance, \citet{2018ApJS..237....2Z}
showed that the spiral-spiral galaxy pairs have higher SFE than spiral-elliptical pairs,
\citet{2022ApJ...934..114Y} found higher SFE during the galaxy-galaxy pericentric passage.

As shown in Fig.~\ref{fig:sfe}, with and without AGN, the pair systems do not deviate significantly from the xGASS sample.
For Pair B without HI detection, 
the lower limit of SFE is already $\sim 1\sigma$ ($\sim$0.5\,dex) higher than the average value of the xGASS detected sample. 
Considering that the molecular gas is also associated with ongoing star formation activities \citep[e.g.][]{2019ApJ...884L..33L},
future CO observations would be useful to confirm the star formation efficiency and possibly the location of the star formation activities.
On the other hand, the other three SFG pairs (A, D, F), lie about 0.3-0.6 dex below the average.
This agrees with a scenario that during the galaxy merger,
the atomic gas fraction enhancement is more significant compared to the SFR enhancement. 
A similar phenomenon is also observed for several post-merger samples \citep[e.g.,][]{2018MNRAS.478.3447E}, but we need a larger sample to confirm this.

\section{Summary}
\label{sec:summary}
This paper presents case studies of comparing optical IFU and single-dish radio telescope 
observations between the H$\alpha$ and \ion{H}{1} emission line profiles in galaxy mergers.
We use the FAST telescope to observe the \ion{H}{1} for a small sample of six galaxy pairs
at different merger stages and of different nuclear activities.
Five systems are detected with secure \ion{H}{1} emissions.
Their \ion{H}{1} line profiles all show irregular shapes, such as broadening, asymmetric peaks, and absorption features,
which is consistent with archival \ion{H}{1} observations on galaxy mergers \citep[e.g.][]{2022ApJ...929...15Z},
and addresses mergers' strong disturbances on the atomic gas.
We do not find galaxy pairs at any stage showing two separable \ion{H}{1} emission lines in the spectra.

We construct the global H$\alpha$ line profiles from MaNGA IFU data and compare them with the \ion{H}{1} line profiles, as a direct approach to compare the ionized gas and the atomic gas.
In summary, the line widths and line profile shapes of \ion{H}{1} and H$\alpha$ are different in all the five \ion{H}{1}-detected galaxy pairs,
suggesting that the disturbance of galaxy merger may have different impacts on the atomic and ionized gases, resulting in unique gas distributions.
Along the line-of-sight velocities, however, some peaks or broadening features in the two profiles can match with each other,
which is indirect evidence that the ionized and atomic gas have correlation in at least certain regions.
The line-of-sight velocity offsets
between the \ion{H}{1} and H$\alpha$ emission line 
centers ($\Delta\,{v_{\rm c50}}$),
shows a decreasing trend toward later merging stages.
This tendency indicates that the merging process
may contribute to the mixing
and eventually align the atomic and ionized gas contents
to the same velocity centers.
Larger statistical samples are needed
to verify this scenario.

The \ion{H}{1} fraction and atomic gas SFE of these six pair systems cover wide parameter spaces,
and all lie within the $3\sigma$ scatter of the large sample results.
We do not find evidence that merger or AGN activities have significant impact on the amount of galaxies' atomic gas.

Compared to the radio interferometry observations, combining optical IFUs and single-dish radio telescope is an efficient way
to study the relative properties of the ionized and atomic gas in galaxies,
but this method was only applied in a few works \citep[e.g.][]{2006ApJS..166..505A,2009ApJ...700.1626A,2023MNRAS.519.1452W}.
We expect new, larger statistical samples with both optical IFU and \ion{H}{1} observations
to study the size, radial distribution, and rotation angle of the atomic gas,
and the method presented in this work could serve as 
an alternative way to analyze the kinematics and distribution
of gas contents of different origins and scales in complex systems 
like the mergers and pairs.

\begin{acknowledgments}
The authors thank the anonymous referees for their constructive comments and suggestions.
We would like to thank Ningyu Tang, Yingjie Jing for their suggestions and help on FAST data reduction. 
We also thank the MaNGA DAP team and NSA team for providing the datacubes and catalogs.
This work is sponsored by the National Key R\&D Program of China (MOST) for grant No.\ 2022YFA1605300, 
the National Nature Science Foundation of China (NSFC) grants No.\ 12273051 and \ 11933003. 
Support for this work is also partly provided by the CASSACA.
This work made use of the data from FAST
(Five-hundred-meter Aperture Spherical radio Telescope). FAST is a Chinese national mega-science facility, operated by National Astronomical Observatories, Chinese Academy of Sciences.
Funding for the Sloan Digital Sky Survey IV has been provided by the Alfred P. Sloan Foundation, the U.S. Department of Energy Office of Science, and the Participating Institutions. SDSS-IV acknowledges
support and resources from the Center for High-Performance Computing at
the University of Utah. The SDSS website is \url{www.sdss.org}.
SDSS-IV is managed by the Astrophysical Research Consortium for the 
Participating Institutions of the SDSS Collaboration including the 
Brazilian Participation Group, the Carnegie Institution for Science, 
Carnegie Mellon University, the Chilean Participation Group, the French Participation Group, Harvard-Smithsonian Center for Astrophysics, 
Instituto de Astrof\'isica de Canarias, The Johns Hopkins University, Kavli Institute for the Physics and Mathematics of the Universe (IPMU) / 
University of Tokyo, the Korean Participation Group, Lawrence Berkeley National Laboratory, 
Leibniz Institut f\"ur Astrophysik Potsdam (AIP),  
Max-Planck-Institut f\"ur Astronomie (MPIA Heidelberg), 
Max-Planck-Institut f\"ur Astrophysik (MPA Garching), 
Max-Planck-Institut f\"ur Extraterrestrische Physik (MPE), 
National Astronomical Observatories of China, New Mexico State University, 
New York University, University of Notre Dame, 
Observat\'ario Nacional / MCTI, The Ohio State University, 
Pennsylvania State University, Shanghai Astronomical Observatory, 
United Kingdom Participation Group,
Universidad Nacional Aut\'onoma de M\'exico, University of Arizona, 
University of Colorado Boulder, University of Oxford, University of Portsmouth, 
University of Utah, University of Virginia, University of Washington, University of Wisconsin, 
Vanderbilt University, and Yale University.

\end{acknowledgments}
\vspace{5mm}
\facilities{FAST \citep{2019SCPMA..6259502J}, SDSS\,2.5\,m \citep{2006AJ....131.2332G}}

\software{Astropy \citep{2013A&A...558A..33A,2018AJ....156..123A}
          }

\appendix
\setcounter{figure}{0}
\renewcommand{\thefigure}{A\arabic{figure}}
\section{FAST data reduction and calibration for On/OFF mode}
\label{ap:dr}

\begin{figure*}
   \gridline{\fig{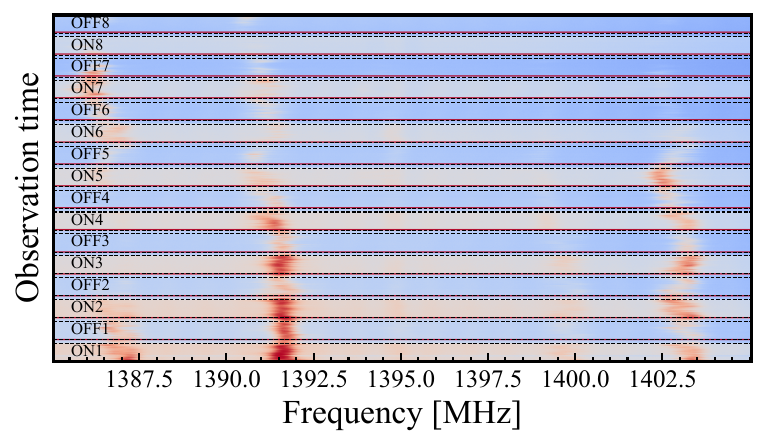}{0.6\textwidth}{(a)}}
   \gridline{\fig{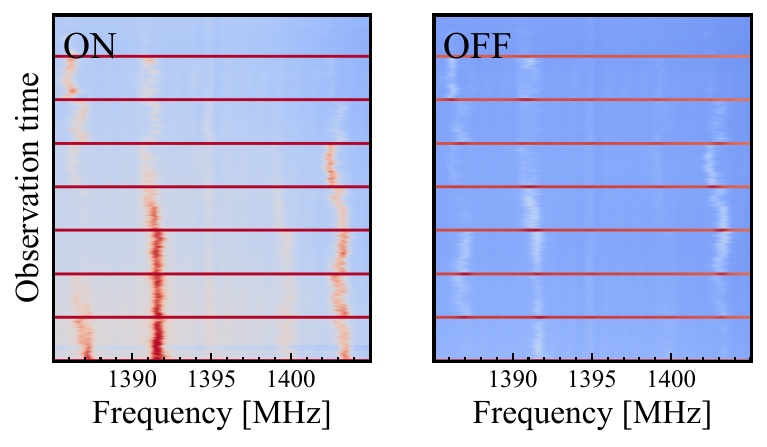}{0.6\textwidth}{(b)}}
   \gridline{\fig{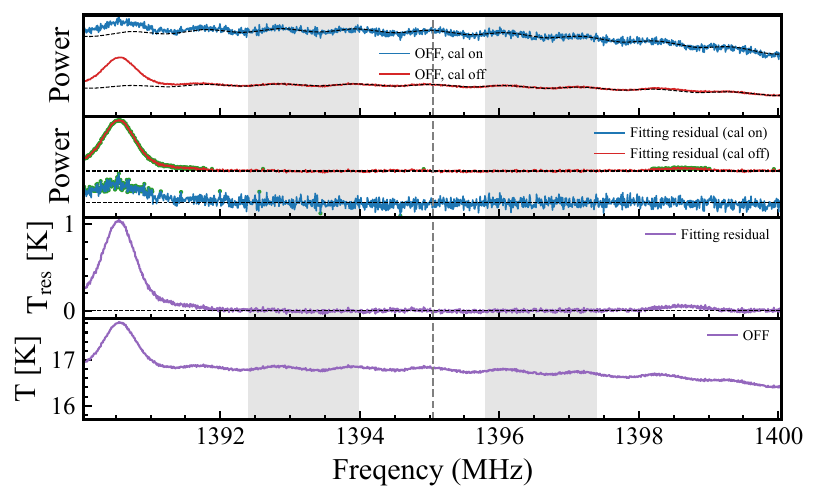}{0.7\textwidth}{(c)}}
   \end{figure*}
   
   \begin{figure*}
   \gridline{\fig{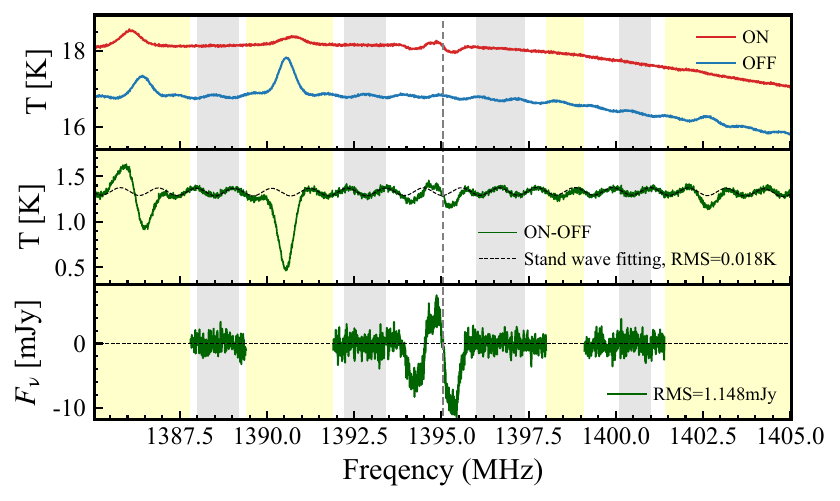}{0.8\textwidth}{(d)}}
   \gridline{\fig{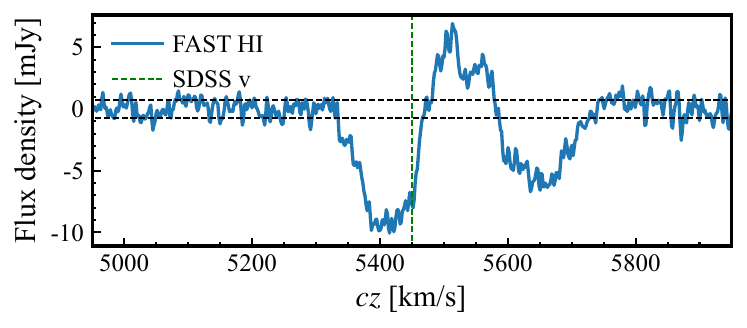}{0.8\textwidth}{(e)}}
   
   \caption{The data reduction procedure adopted in this work, using the data from Pair E as an illustration.
   {\bf Step 1 (a):} The `waterfall' map of the raw data in a 16\,MHz frequency range. The crimson signals are the RFIs and calibration noises.
   {\bf Step 2 (b):} The `waterfall' maps of combined ON or OFF observations after removing the time domain RFIs (if exist).
   {\bf Step 3 (c):} Example temperature calibration of a 5-minutes OFF-target observation. 
   The gray backgrounds are the selected frequency range with no spectra features used for baseline fitting,
    and the gray dashed line in the center marks the expected \ion{H}{1} frequency from the optical redshift.
   Panel 1 and 2 are the noise-on (red) and noise-off (blue) spectra
   before and after baseline (black curves in panel 1) removal. 
   Green dots in panel 2 mark the possible RFI frequencies.  
   Black lines in panel 2 and 3 mark the zero point baseline.
   As a check, panel 3 shows the resulting temperature profile (purple) combining the noise-on and noise-off observations.
   Panel 4 is the combined noise-off and noise-on spectra to be used in the next step,
   after converting the counts into temperatures. 
   {\bf Step 4 (d):} Get the \ion{H}{1} spectra (lower panel) from each set of neighboring ON and OFF observations derived in (c).
   Baseline is fitted (black dashed lines) and removed.
   Yellow strips mask the frequency range heavily affected by the RFI,
   and these frequency regions are not used for baseline fitting.
   {\bf Step 5 (e):} Stacking all \ion{H}{1} spectra derived in (d) as the final result. 
   The weight of each exposure is calculated by the rms in the featureless gray zones.
      \label{fig:pipeline}}
   \end{figure*}

We adopt and modify the data reduction and calibration procedure from \citet{2020RAA....20...64J} and \citet{2020A&A...638L..14C}. 
We summarize and illustrate the procedure in Fig.~\ref{fig:pipeline},
and is described as follows:

{\bf 1. Remove the time-domain RFIs:}
We find that the RFIs can be divided into two types,
the time-domain RFIs (hereafter t-RFIs) and the frequency-domain RFIs (hereafter f-RFIs).
Usually, the t-RFIs are strong and will appear or disappear relatively abruptly.
The f-RFIs usually appear continuously in a narrow frequency range of several MHz,
and are weaker than the t-RFIs,
thus we will remove the f-RFIs by masking relative frequencies later in step 3.
In the first step, we manually identify and remove the t-RFIs by their strong fluxes, 
as shown in Fig.~\ref{fig:pipeline}(a).
In this way, we get the ON/OFF spectra not affected by t-RFIs, as shown in Fig.~\ref{fig:pipeline}(b).

{\bf 2. Temperature calibration:}
As mentioned in Sec~\ref{subsec:setting},
the calibration noise is injected during the first 20 seconds of each ON and each OFF exposure.
For each 16 MHz frequency bandpass, the antenna temperature profile is calculated as:
\begin{equation}
\rm
T_{noise,off}(\nu) = T_{noise,inj}(\nu)\times
\frac{f_{noise,off}(\nu)}{\overline{f
_{noise,on}}(\nu)-\overline{f_{noise,off}}(\nu)}
,\label{eq:cal}
\end{equation}
where $\rm T_{noise,inj}(\nu)$ is the injected noise temperature in Kelvin, 
$\rm T_{noise,off}$ is temperature with no injected noise, sometimes referred to the antenna Temperature, 
$\rm f_{noise,off}(\nu)$ is the mean digital counts strengths as a function of frequency in the noise off mode,
and
$\rm \overline{f_*}$ is the mean digital counts strengths in the selected frequency ranges, 
for noise on and noise off, respectively.
In our observations, 
we use the noise diode noise temperature profile, 
which was reported to have a temperature varying around 5.4K for the 
adopted `high power' injected calibration noise at the time of observation
\citep[see Section 3.1 of][]{2020RAA....20...64J}.
The choice of averaging the $\rm \overline{f_*}$ is
to reduce the influences of f-RFIs and signals on the temperature calibration.
In this step, we carefully avoid the regions with 
possible f-RFI signals (as marked in green in Fig.~\ref{fig:pipeline}(c), panel 3, see below). 

In our calibration frequency range of $\sim$16MHz,
the baseline is dominated by the stable standing wave with a $\sim$1MHz period,
thus can be fitted by a polynomial plus a sinusoidal function.
As shown in Fig.~\ref{fig:pipeline}(c), we calculate 
the $\rm \overline{f}(\nu)$ by fitting a (polynomial+sinusoidal) baseline profile
to the spectra in the frequency ranges (gray stripes) 
that are least affected by signals or f-RFIs.
The top two panels show the spectra for noise on (blue) and noise off (red) modes
before and after removing the fitted baseline;
while the bottom two panels show the resulting $\rm T_{noise,off}(\nu)$
after and before removing the fitted baseline profile. 
In addition, if the OFF spectra in noise-off mode at the expected \ion{H}{1} frequency
still show residual values after the baseline removal,
we mask the exposure as `failed calibration'
and exclude it from the stacking in the next step, 
to make sure that OFF observations can serve as `no flux' reference.
This only applies to a few exposures. 

{\bf 3. Obtain the spectra from each ON observation:}
As shown in Fig.~\ref{fig:pipeline}(d), we select the contiguous OFF spectrum (red, $\rm T_{OFF}$)
as the baseline for each ON observation (blue, $\rm T_{ON}$).
Then we follow Equation \ref{eq:hical}
to obtain the spectra (green spectrum in middle panel) for baseline fitting and f-RFI removing.
\begin{equation}
\rm
T_{HI}(\nu) =T_{OFF}(\nu)\times
\frac{T_{ON}(\nu)-T_{OFF}(\nu)}{\overline{T_{OFF}}}
,\label{eq:hical}
\end{equation}
Middle panel of Fig.~\ref{fig:pipeline}(d) shows the ON-OFF spectra, 
where the yellow strips are the masked out f-RFI frequency regions,
the light gray strips are the featureless regions used for baseline fitting.
The standing wave baseline is then fitted in these featureless regions using a polynomial+sinusoidal profile,
as plotted in black lines in the middle panel.
The standing wave is caused by the reflection between the receiver and the mirror panels,
thus has a period of $\sim$1MHz ($\sim$300\,meter$/c$).
For reliability, we make sure that the baseline fitting frequency regions span at least $\sim$4MHz ($>$4 sinusoidal periods). 
The green spectrum in the lower panel is the residual spectra after removing the fitted baseline
and masking out f-RFI regions, 
i.e. the \ion{H}{1} spectrum for one set of ON observations. 
The temperature are then converted to flux density using 
the gain profile at the targeted frequencies \citep{2020RAA....20...64J}, 
which varies around 16.0\,K\,$\rm Jy^{-1}$ for all of our targets.
When the zenith angle of the source is higher than 26.4\textdegree, 
the effective aperture of FAST become smaller, resulting in a lower gain.
We correct the gain based on the test in \citet{2020RAA....20...64J} in this step.
We derive the spectrum for each ON exposure first, 
because during each 40 minutes observations,
the system temperature and the gain of FAST receiver varies.
The rms are calculated from the residual in the featureless gray zones. 

{\bf 4. Spectra stacking and \ion{H}{1} mass:}
After we obtain the ON-OFF spectra of all the ON exposures for the same target,
we stack the baseline-removed spectra, weighted by their rms.
Then the spectra are Hanning-smoothed to a velocity resolution of $\sim$3.3\,km\,$s^{-1}$ 
as our final results.
The \ion{H}{1} spectra of the six pairs are shown in Fig.~\ref{fig:all}.
We find five \ion{H}{1} detections with high S/N and 
one non-detection with low rms to restrict the upper limit of \ion{H}{1} mass (Pair B).
Green dashed line(s) show the optical redshift(s) of the member galaxies,
and red dashed line corresponds to the center of the detected \ion{H}{1} emission line.

\setcounter{figure}{0}
\renewcommand{\thefigure}{B\arabic{figure}}

\section{The Sensitivity of FAST: Compared to GBT and Arecibo}\label{ap:sensitivity}
The sensitivity of a single-dish telescope is proportionate to the square of the aperture diameter
and also depends on the system temperature.
Here we compare the sensitivity of our FAST observation with the GBT and Arecibo observations 
when using the ON-OFF mode to observe emission lines.
For GBT observations, we use the result of the \ion{H}{1}-MaNGA survey.
For Arecibo observations, we use the results from the GASS survey.
They both used the ON-OFF observation mode
and share similar data reduction methods with this paper,
thus are suitable for comparing the sensitivity.
For their observation setting-ups and data reduction, we refer the readers to \citet{2019MNRAS.488.3396M} and \citet{2010MNRAS.403..683C}, respectively.

\begin{figure*}
   \plotone{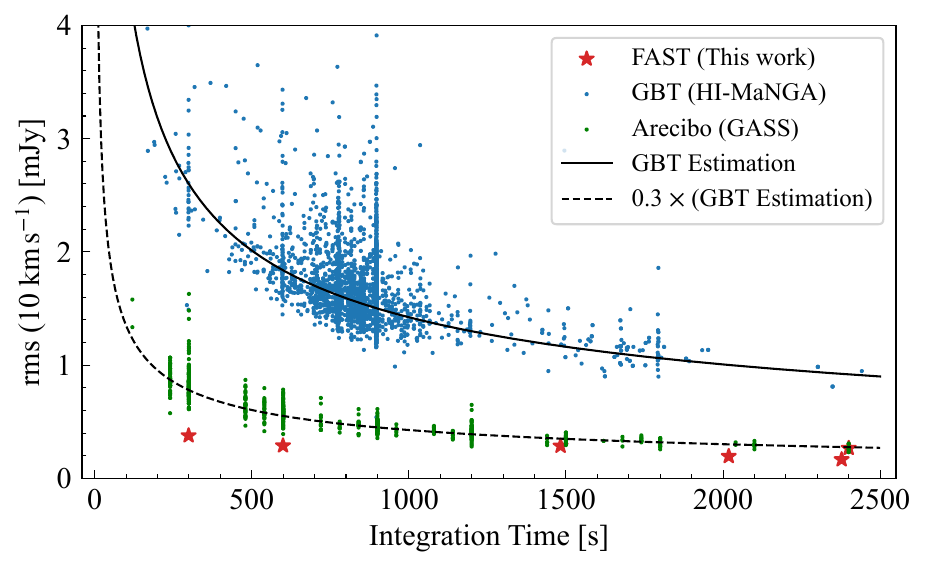}
   \caption{The rms versus integration times from HI-MaNGA survey (blue dots), GASS survey (green dots), and our FAST observations (red stars). All the rms value are corrected to the velocity resolution of 10\,$\rm km\,s^{-1}$.}
   \label{fig:rms}
   \end{figure*}

Here we use the rms when the velocity resolution is converted to $\rm 10\,km\,s^{-1}$
to compare the sensitivity of different telescopes.
Ideally, the rms of an observation can be expressed as:
$\rm rms \propto t^{-0.5} \times dv^{0.5}$.
At given integration time and velocity resolution,
lower rms means higher sensitivity.
In Fig.~\ref{fig:rms}, we plot the rms versus integration time of the observations from different telescopes.
The blue dots, green dots and red stars are the GBT, Arecibo, and FAST observations, respectively.
The black line is the relation used for estimating the rms of a GBT observation.
We also plot the 0.3$\times$ of the relation as the black-dashed line to guide the eyes for comparison.
For shorter time observations, the sensitivity of FAST is $\sim 5 \times$ better than that of GBT,
and $\sim 2 \times$ than that of Arecibo.
For long-time exposure, however, the rms of FAST observation does not significantly decrease following the theoretical predication.
This phenomenon was also found in \citet{2020A&A...638L..14C}.
One possible reason is found during our data reduction,
that the RFI can increase the system temperature, resulting in higher rms.
This is the performance of FAST at the date of early 2021.
Future improvements are needed for FAST to suppress the rms for long-time spectral line observation mode.

\bibliography{article}{}
\bibliographystyle{aasjournal}

\end{document}